\documentstyle[aps,prb,epsf,rotate,multicol]{revtex}

\begin{document}

\author{D. Rainer$^a$, J. A. Sauls$^b$ and D. Waxman$^c$}

\address{$^a$Physikalisches Institut,
     Universit\"at Bayreuth,
     D-95440 Bayreuth,
     Germany \\
$^b$Department of Physics and Astronomy, Northwestern University,
Evanston, Illinois, USA \\
$^c$School of Mathematical and Physical Sciences,
The University of Sussex, Brighton BN1 9QH, Sussex, England}

\title{The Current Carried by Bound States of a Superconducting Vortex}
%\date{October 15, 1995}
\date{submitted April 17, 1996}
\maketitle

\begin{abstract}
We calculate the spectrum of quasiparticle excitations in the core of
isolated pancake vortices in clean layered superconductors. We show
that both the circular current around the vortex center as well as any
transport current through the vortex  core is carried by localized
states  bound to the core by Andreev scattering.  Hence the physical
properties of the core  are governed in clean high-$\kappa$
superconductors (e.g. the cuprate superconductors) by the Andreev bound
states,  and not by normal  electrons as it is the case for traditional
(dirty) high-$\kappa$ superconductors.
\end{abstract}

\section{Introduction}

We discuss specific aspects of the core of a vortex line in layered
high T$_c$ superconductors. The physics of these vortices is governed
by two distinct length scales, the London penetration depth in the
planes, $\lambda _{\Vert }\approx 10^3$\AA , and the coherence length
in the planes, $\xi _{\Vert }\approx 10-20$\AA . The penetration depth
is the electromagnetic length scale of a vortex. The physics on this
length scale is well described by a combination of macroscopic
electromagnetism, London's theory for supercurrents along the layers,
and interlayer Josephson coupling. This  description breaks down in the
core of the vortex, {\it i.e.} at distances of order $\xi_{\Vert}$ from
the center of the vortex. Thus, physical properties of the core carry
information on the microscopic physics of high T$_c$
superconductivity.  The small coherence length of high T$_c$
superconductors makes the vortex core a good potential sensor for
microscopic mechanisms of superconductivity.  Our discussion of the
vortex core in high T$_c$ superconductors is based on the Fermi-liquid
model of superconductivity. The physical properties of the vortex core
predicted by this model are spectacular, unique, and could serve as
fingerprints of the traditional pairing theory of superconductivity.

The vortex core of traditional high-$\kappa $ superconductors is well
described by the Bardeen-Stephen model\cite{bar65} which represents the
core by a region of normal electrons. The Bardeen-Stephen model is
justified as long as the mean free path, $\ell $, is much shorter than
the core size, so that the motion of an electron gets randomized before
it leaves the core. This condition is not fulfilled in high T$_c$
superconductors which are generally clean superconductors with $\ell
>\xi _{\Vert }$. The core of a vortex in a clean superconductor was
first studied in the classic papers of Caroli, Matricon, and de
Gennes.\cite{car64,car65} These authors calculated the spectrum of
quasiparticle states in the core, and showed that electrons and holes
form bound states at energies below the bulk energy gap. Further early
studies of the excitation spectrum in the core can be found in Refs.
\onlinecite{bar69,kra74}.

More recent theoretical work was stimulated by the direct observation
of core states in $NbSe_2$ by scanning tunnelling
spectroscopy (STS).\cite{hes89f,ren91} The recent report of STS in YBCO
\cite{ren95,mag95} provides new information on the excitation spectrum
of vortices in the high $T_c$ cuprates. Consequently, theoretical
efforts focused on the tunnelling density of states of bound states in
isolated vortices and vortex
lattices.\cite{sho89,kle89,kle90,gyg90,gyg90a,ull90,poe93} These
calculations show that the bound states in the core have a different
nature compared with the usual quantum mechanical bound states in a
potential well. The core states are coherent superpositions of particle
states and hole states and are formed by repeated Andreev scattering
from the pair potential (order parameter) in the core.  Andreev
scattering is a process of ``retroreflection'' of excitations:  spatial
variations of the amplitude or the phase of the order parameter induce
branch conversion of electron-like excitations into hole-like
excitations, and vice versa. Bound states occur at energies at which
the phases of multiply reflected electron-like and hole-like states
interfere constructively. The charge current carried by an incoming
electron and an outgoing Andreev reflected  hole is identical because
the reversal of the velocity in an Andreev reflection process is
compensated by the reversal of the charge due to electron-hole
conversion. Consequently, Andreev bound states can transport a charge
current, unlike bound states in a potential well. Charge conservation
requires that the current carried by the bound states inside the core
is transported outside the core by bulk supercurrents.  This leads to
an interplay between supercurrents flowing past the core and the bound
states in the core. Hence, the physics of the `normal core' in clean
superconductors is basically the physics of the bound states in contact
and intimate exchange with the superconducting environment outside the core.

Consider a stack of ``pancake'' vortices forming an isolated vortex
line whose axis is oriented perpendicular to the layers. We investigate
the current distribution in the core of a pancake vortex, and show how
this distribution changes if the vortex is exposed to a bulk
supercurrent, or the circulation is changed from  $2\pi$ to $4\pi$. We
calculate the {\it spectral current density}, which carries the
information on the contribution of the states in a given energy
interval to the total current density. A supercurrent in homogeneous
superconductors is distributed over all continuum states. These states
exhibit Doppler shifts of their energies, $\delta \epsilon ={\bf
v}_f\cdot {\bf p}_s$, in the presence of a phase gradient in the order
parameter (or superfluid momentum, ${\bf p}_s = \case{\hbar}{2}
\mbox{\boldmath $\nabla$}\chi -\case{e}{c}{\bf A}$). The total current
is obtained by adding the contributions of states with positive shifts
from quasiparticles co-moving with the flow and the contributions with
negative shifts from quasiparticles that are counter-moving relative to
the flow field. We find that the currents in the core have a very
different spectral distribution from bulk supercurrents.  The continuum
states (scattering states) show smeared out Doppler shifts, and
contribute very little to the total current. The dominant contributions
to the circulating currents around the vortex center, as well as the
currents through the core, come from Andreev bound states.  Hence, the
physics of vortex cores in clean superconductors ($\xi_{\Vert} \ll \ell
$) is very different from the physics of the vortex core in a dirty
superconductor ($\ell \ll \xi_{\Vert} $), which is well described by a
continuum of normal electronic states. The calculations presented in
this paper concentrate on stationary properties of the vortex core of
clean layered superconductors. We expect more spectacular effects in
the dynamic properties. The bound states react sensitively to the
environment outside of the core. This leads to a coupling of the
collective degrees of freedom in the London range of the vortex and the
bound states in the core, which will produce a rich spectrum of largely
unexplored dynamical phenomena.

Below we present analytical and numerical  calculations for the states
in the vortex core. We use two versions of a quasiclassical formulation
of the BCS theory of superconductivity: a) {\sl Andreev's
theory}\cite{and64} which represents the quasi\-classical limit of
Bogolyubov's equations,\cite{bog58} and b) the {\sl quasiclassical
theory} of Eilenberger \cite{eil68}, Larkin, and
Ovchinnikov\cite{lar69} which represents the quasiclassical limit of
Gorkov's Green's function theory. Andreev's theory and the
quasiclassical theory are essentially equivalent for clean
superconductors, and in this limit the choice of approach is largely a
matter of taste. However, the quasiclassical theory has a wider range
of application. It is the generalization of Landau's Fermi liquid
theory to the superconducting state, and is capable of describing a
broader range of superconducting materials and phenomena, such as dirty
superconductors or superconductors with short inelastic lifetimes
(strong-coupling superconductors).\cite{rai94}  Section II  contains
analytical results for the bound states and the spectral current
density for a pancake vortex with a superimposed bulk supercurrent.
These results are obtained from Andreev's Hamiltonian\cite{and64} by
the methods described in Ref. \onlinecite{wax93}.  In section III we
discuss the numerical results, which are obtained using the
quasiclassical theory of Fermi-liquid superconductivity. We solve the
quasiclassical transport equations to obtain self-consistently the pair
amplitude (order parameter) for pancake vortices.  Given the pair
amplitude  we calculate the excitation spectrum in the core of the
vortex, and deduce from it the spectral current density. The numerical
calculations are done for layered superconductors with s-wave pairing.
The analytical  and numerical results  confirm and complement each
other, and they establish the important role of the bound states for
the currents in the  core region of a vortex.

\section{Analytical Results}\label{analyt}

In this section, we  investigate the spectrum of current carrying
states of a two-dimensional pancake vortex in equilibrium at
temperature $T$, in the presence of an externally imposed
supercurrent. We ignore  the spin degree of freedom of a
quasiparticle  excitation;\cite{f1} in this case it is sufficient
to work in the two-dimensional space of particle-hole degrees
of freedom. Operators in this space are $2\times 2$ matrices, and we
use the notation $\hat\tau_{1}$, $\hat\tau_{2}$, $\hat\tau_{3}$, for
the three Pauli matrices in particle-hole space (Nambu space), and
$\hat 1$ for the unit matrix.  The Hamiltonian for the quasiparticle
excitations, \cite{deg66} the Bogolyubov Hamiltonian, then reads:
\begin{equation}
\label{bogoliubov}
\hat{H}_B=h_0(\hat{{\bf p}}+\frac{e}{c}{\bf A(}\hat{{\bf  
r}}))\hat\tau_{3}+\hat\Delta (\hat{{\bf r}}),\qquad h_0({\bf p})=\left(
\frac{  {\bf p}^2-p_f^2}{2m}\right) \,,
\end{equation}
where $\hat{{\bf p}}=(\hat{p}_x,\hat{p}_y)$ and $\hat{{\bf r}  }{\bf
\ =(}\hat{x},\hat{y}{\bf )}$ are the momentum and position operators
appropriate to a particle moving in two dimensions, $p_f\equiv mv_f$ is
the Fermi momentum and ${\bf A}$ is the electromagnetic vector
potential. The order parameter, $\hat\Delta ({\bf r  })$ is an
off-diagonal matrix and is generally represented by a linear
combination of $\hat\tau_{1}$ and $\hat\tau_{2}$.

In the absence of an externally imposed supercurrent, we write the
order parameter of the vortex as
\begin{equation}
\label{delta0}\Delta_0({\bf r})=\Delta_{0}\text{\/}F(r)\hat\tau_{1}\exp
(i\varphi\,\hat\tau_{3}) \,,
\end{equation}
where $\Delta_{0}$ is the magnitude of the order-parameter of a bulk
superconductor at temperature $T$, $F(r)$ is the normalized profile of
the vortex,  which is a monotonically increasing function of $r$
obeying $F(0)=0,\ F(\infty )=1$, and $\varphi $ is the angular
coordinate of $  {\bf r}$ with $x=r\cos \varphi $, $y=r\sin \varphi .$
The main assumption we make in this section is that the order-parameter
in the presence of a superflow ${\bf p}_s = \hbar/2\,\mbox{\boldmath
$\nabla$}\chi - e/c\,{\bf A}$ has the form
\begin{equation}
\label{delta}
\hat\Delta({\bf r})=
\exp (+\frac{i}{2}\mbox{\boldmath $\nabla$}\chi\cdot{\bf r}\,\hat\tau_{3})
\hat\Delta_0({\bf r}) 
\exp (-\frac{i}{2}\mbox{\boldmath $\nabla$}\chi\cdot{\bf r}\,\hat\tau_{3})
\end{equation}
We assume throughout this section that $p_s$ is small compared to the
bulk critical current, $v_f p_s \ll \Delta_{0}$.

The principal physical quantities with which we shall concern ourselves are
the {\it spectral current density}, and the total equilibrium current density
which is related to ${\bf j}({\bf r},\epsilon)$ by
\begin{equation}
{\bf j}({\bf r},T)=\int d\epsilon \ {\bf j}({\bf r} ,\epsilon)\/f(\epsilon
),\qquad f(\epsilon )=\frac 1{\exp (\epsilon /T)+1}
\,. 
\end{equation}
We shall also make reference to the local density of states, 
$N ({\bf r} ,\epsilon)$. 
The quantities ${\bf j}({\bf r} ,\epsilon)$ and 
$N ({\bf r},\epsilon)$ may be expressed in terms of the 
one-particle Greens function, $ 
1/(\epsilon -\hat{H}_B)$ or, equivalently, the ``spectral function'' $ 
\delta (\epsilon -\hat{H}_B)$. Using the spectral function, we find that
in Dirac notation, 
\begin{equation}
{\bf j}({\bf r} ,\epsilon) = 2e\langle 
{\bf r|}\{\frac{\hat{{\bf p}}+\frac{e}{c}{\bf A}}{2m},\delta (\epsilon -\hat{H 
}_B)\}|{\bf r\rangle }_{1,1} \,,
\end{equation}
\begin{equation}
N ({\bf r} ,\epsilon) = 2\langle {\bf r|}\delta (\epsilon -\hat{H} 
_B)|{\bf r\rangle }_{1,1} \,,
\end{equation}
where the subscript $1,1$ denotes the upper left element of the
$2\times 2$ matrices, thereby selecting out the {\it particle sector}
of the spectral function and the factor $2$ takes into account both
spin projections of the quasiparticles.

\subsection{Andreev Hamiltonian}

Most calculations of the properties of superconductors with
inhomogeneous order parameters are simpler in the quasiclassical limit,
where one takes advantage of the separation in the scales of the
wavelength of quasiparticles near the Fermi energy and the
characteristic scale for spatial variations of the pair potential,
i.e.  $\hbar/p_f\ll\xi_0$. The quasiclassical limit of the Bogolyubov
Hamiltonian (\ref{bogoliubov}) is the Andreev Hamiltonian in which the
kinetic energy in (\ref{bogoliubov}) is replaced by an operator that is
{\it linear} in the gradient.\cite{and64} Let us define the
normal-state density of states at the Fermi level, $N_f=p_f/2\pi v_f$,
and introduce the directions, $\hat{\bf k}=(\cos \varphi _k,\sin
\varphi _k)$ and $\hat{\bf l}=(-\sin \varphi _k,\cos \varphi _k)$,
which are, respectively, parallel and perpendicular to trajectories of
a quasiparticle wavepacket in the quasiclassical description, {\it
i.e.} ${\bf v}_f=v_f\hat{\bf k}$. The coordinates along these
directions are defined by ${\bf r}=\zeta \hat{\bf k}+\eta \hat{\bf
l}$.  In addition we work in the limit $\lambda/\xi\gg 1$,
in which case the vector potential is approximately constant in the
vicinity of the vortex core and can be neglected. The order parameter
in (\ref{delta}) can be written as
\begin{equation}
\label{op}
\hat \Delta ({\bf r})=\Delta_{0}F(r)\,
\hat{U}\,
\frac{(\hat\tau_{1}\zeta +\hat\tau_{2}\eta )}{r}\,
\hat{U}^{\dagger}\,,
\qquad
\hat{U}=\exp \left[ +i{\bf p}_s\cdot\hat{\bf k}\zeta\hat\tau_{3}\right] . 
\end{equation}
By performing a gauge transformation that removes the factor of
$\hat{U}$ in Eq. (\ref{op}) we obtain the spectral current density and
the local density of states in terms of the Andreev Hamiltonian for an
isolated vortex,
\begin{equation}
\label{andreev}
\hat{H}_A=v_f\/\hat{p}_\zeta \hat\tau_{3}+\Delta_{0} 
\frac{F(\hat{r})}{\hat{r}}
(\hat\tau_{1}\hat{\zeta }+\hat\tau_{2}\eta
), 
\end{equation}
\begin{equation}
{\bf j}({\bf r} ,\epsilon) \simeq  4\pi e v_f^2N_f\int_0^{2\pi }
\frac{d\varphi _k}{2\pi }\text{\/\thinspace }\hat{\bf k}\text{\/\thinspace } 
\langle \zeta |\delta (\epsilon -[\hat{H}_A+v_f{\bf p}_s\cdot\hat{\bf k} 
])|\zeta \rangle _{1,1} \,,
\end{equation}
\begin{equation}
N ({\bf r} ,\epsilon) \simeq  4\pi v_fN_f\int_0^{2\pi }\frac{ 
d\varphi _k}{2\pi }\langle \zeta |\delta (\epsilon -
[\hat{H}_A+v_f{\bf p}_s\cdot\hat{\bf k}])|\zeta \rangle _{1,1}\,,
\end{equation}
where $|\zeta\rangle$ is an eigenvector of the ``one-dimensional''
trajectory coordinate operator, $\hat{\zeta }$:  $\hat{\zeta }|\zeta
\rangle =\zeta |\zeta \rangle$.  The operators $\hat{\zeta}$ and
$\hat{\bf k}\cdot\hat{\bf p}=\hat{p}_\zeta$ appearing in $\hat{H}_A$
are canonically conjugate: $[\hat{\zeta },\hat{p}_\zeta ]=i\hbar$.
The quasiclassical interpretation given to (\ref{andreev}) is as
follows: quantum-mechanical evolution in particle-hole space takes
place along classical trajectories parallel to $\hat{\bf k}$ having a
fixed value of $\eta \equiv \hat{\bf l}\cdot{\bf r}$. Thus, $\eta$ is
identified as a c-number impact parameter.\cite{f3}

\subsection{\label{considerations}The current density of a vortex in a flow field}

Let us write the current density at temperature $T$ as 
\begin{equation}
{\bf j}({\bf r,}T)=\int_{-\Lambda }^\infty d\epsilon \ {\bf j}(\epsilon , 
{\bf r})\text{\/}f(\epsilon ), 
\end{equation}
where $\Lambda$ is a high energy cutoff that serves to make manipulations
of ${\bf j}({\bf r,}T)$ well defined; large positive energies are
automatically cut off by the Fermi function, $f(\epsilon )$. Where no
ambiguity arises, we shall take $\Lambda =\infty$. Defining 
$\label{q}{\bf q}=v_f{\bf p}_s$, and using (\ref{andreev}) we have 
\begin{equation}
\label{joft}
{\bf j}({\bf r,}T) = 4\pi e v_f^2N_f\int_0^{2\pi }\frac{d\varphi
_k}{2\pi }\int_{-\Lambda -{\bf q}\cdot\hat{\bf k}}^\infty d\epsilon\,
\text{\/}\hat{\bf k}\text{\/}\langle \zeta |\delta (\epsilon
-\hat{H}_A)|\zeta \rangle _{1,1}\text{\/} f(\epsilon +{\bf
q\cdot}\hat{\bf k})
\,. 
\end{equation}
Next, we split up the energy integrals into the following terms 
\begin{equation}
{\bf j}({\bf r,}T) = {\bf j}_1({\bf r,}T)+{\bf j}_2({\bf r,}T)+{\bf j}_3( 
{\bf r,}T) \,,
\end{equation}
\begin{equation}
{\bf j}_1({\bf r,}T) =  4\pi ev_f^2N_f\int_0^{2\pi } 
\frac{d\varphi _k}{2\pi }\text{\/}\hat{\bf k}\text{\/}\left( {\bf q\cdot} 
\hat{\bf k}\right) \langle \zeta |\delta (-\Lambda -\hat{H}_A)|\zeta
\rangle _{1,1} \,,
\end{equation}
\begin{equation}
{\bf j}_2({\bf r,}T) = 4\pi ev_f^2N_f\int_0^{2\pi } 
\frac{d\varphi _k}{2\pi }\int_{-\infty }^\infty d\epsilon \text{\/} 
\,\hat{\bf k}\text{\/}\langle \zeta |\delta (\epsilon -\hat{H}_A)|\zeta
\rangle _{1,1}\left[ \text{\/}f(\epsilon +{\bf q\cdot}\hat{\bf k})-f(\epsilon
)\right] \,,
\end{equation}
\begin{equation}
{\bf j}_3({\bf r,}T) = 4\pi ev_f^2N_f\int_0^{2\pi }\frac{d\varphi _k}{ 
2\pi }\int_{-\Lambda }^\infty d\epsilon \text{\/}\,\hat{\bf k}\text{\/} 
\langle \zeta |\delta (\epsilon -\hat{H}_A)|\zeta \rangle _{1,1}\text{\/} 
f(\epsilon)\,. 
\end{equation}
The three contributions to the current have different interpretations.
\begin{enumerate}
\item  Since $\Lambda $ is large, $\langle \zeta |\delta (-\Lambda -\hat{ 
H}_A)|\zeta \rangle _{1,1}$ may be replaced by its high energy, normal-state
limit, $1/2\pi v_f$ and 
\begin{equation}
\label{j1}{\bf j}_1({\bf r,}T)=ev_f^2N_f{\bf p}_s.
\end{equation}
This term coincides with the $T=0$ current of a uniform superconductor.
\item  The term ${\bf j}_2({\bf r,}T)$ contributes to
``backflow'', since it always yields a current with a component in the $- 
\hat{{\bf p}}_s$ direction.\cite{f4}
The term ${\bf j}_2({\bf r,}T)$ contains the current carried by the
bound states and also a correction to the $T=0$ current due to the
thermal breaking of pairs. To appreciate these points we look at this
term in two limits, assuming $q\ll \Delta_{0}$.  (i) For $T=0$ we have
\begin{equation}
\label{j2}{\bf j}_2({\bf r,}0)=4\pi ev_f^2N_f\int_0^{2\pi }\frac{d\varphi
_k}{2\pi }\int_0^{-{\bf q\cdot}\hat{\bf k}}d\epsilon \,\text{\/\/}\hat{\bf k} 
\text{\/\thinspace \thinspace }\langle \zeta |\delta (\epsilon -\hat{H} 
_A)|\zeta \rangle _{1,1}.
\end{equation}
The small size of $q$ ensures that the energy integral only selects states in the
gap, thus ${\bf j}_2({\bf r,}0)$ only obtains contributions from the bound
states. (ii) For  $T\neq 0$ and  assuming $\Delta _0({\bf r})$ 
to be that of a uniform system,
$\Delta _0({\bf r})=\Delta_{0}\hat\tau_{1},$ 
we write $f(\epsilon +{\bf q\cdot}\hat{\bf k})-f(\epsilon )\approx 
{\bf q\cdot}\hat{\bf k}$\/$f$\/$^{\prime }(\epsilon )$ and obtain
\begin{equation}
{\bf j}_2({\bf r,}T)$ = $-ev_f^2N_f{\bf p}_s\int_{-\infty }^\infty
d\epsilon $\/$\frac{|\epsilon |}{\sqrt{\epsilon ^2-\Delta_{0}^2}}\Theta
(\epsilon ^2-\Delta_{0}^2)[-f$\/$^{\prime }(\epsilon )]=
-ev_f^2N_f{\bf p}_sY(\beta \Delta_{0})
\,,
\end{equation}
where $Y(\beta \Delta_{0})$ is the Yosida function which gives a
quantitative measure of the thermal breaking of Cooper pairs.  \item
The term ${\bf j}_3({\bf r,}T)$ is independent of ${\bf p}_s$ and is
simply the current density of a vortex in the absence of an externally
imposed supercurrent.
\end{enumerate}

\subsection{Current carrying bound states at the center of the vortex}

Consider the spectral properties at the center of the vortex. For $\eta =0$
\begin{equation}
\label{zeroeta}\left. \hat H_A\right| _{\eta =0}=v_f\text{\/}\hat{p}_\zeta
\hat\tau_{3}+\Delta_{0}\text{\/}F(\hat{\zeta })\hat\tau_{1}\,,
\end{equation}
with $F(-\hat\zeta)=-F(\hat\zeta)$ accounting for the $\pi$ phase
change across the vortex core. This special case of the Andreev
Hamiltonian is identical in form to the continuum Hamiltonian used to
describe {\it trans}-polyacetylene containing a single topological
soliton.\cite{hee88} It is known that this Hamiltonian always has a
non-degenerate bound state at zero energy.\cite{ati75} Whether or not
it has other bound states depends on the form of the profile, $F(\zeta
)$. For the single quantum vortex and trajectories through the center
there are no other bound states. The eigenfunction for the zero-energy
bound state, $\psi _0(\zeta )$, is found by solving $\left[
v_f\frac{\partial _\zeta } i\hat\tau_{3}+\Delta_{0}\/F(\zeta
)\hat\tau_{1}\right] \psi_0(\zeta )=0.$ The normalized solution is
\begin{equation}
\psi _0(\zeta ) = \frac 1{\sqrt{L}}\exp \left( -\frac{\Delta_{0}}{v_f 
}\int_0^\zeta d\zeta ^{\prime }\/F(\zeta ^{\prime })\right) \left( 
\begin{array}{c}
\frac 1{
\sqrt{2}} \\ -\frac i{\sqrt{2}} 
\end{array}
\right) \,,
\end{equation}
\begin{equation}
L = 2\int_0^\infty d\zeta \exp \left( -\frac{2\Delta_{0}}{v_f} 
\int_0^\zeta d\zeta ^{\prime }\/F(\zeta ^{\prime })\right) \,,
\end{equation}
where $L$ is a profile dependent quantity with the dimensions of
length, $L\sim v_f/\Delta_{0}$.  Analytical estimates of the bound
states at distances far from the vortex are given in appendix
\ref{far}.

For energies $|\epsilon |<\Delta_{0}$ only the bound
state of $H_A$ will contribute to the spectral current density (and the
local density of states),
\begin{equation}
\label{jo} 
{\bf j}(\epsilon ,{\bf 0}) \simeq  4\pi e v_f^2 N_f\int_0^{2\pi } 
\frac{d\varphi _k}{2\pi }\,\hat{\bf k}\,\delta (\epsilon -v_f{\bf p}_s\cdot 
\hat{\bf k})\left[ \psi_0(0)\psi _0^{\dagger }(0)\right]_{1,1} 
\end{equation}
\begin{equation}
\qquad = \left[ \frac \epsilon {p_s}\frac{2eN_f\Delta_{0}}{L/\xi
}\frac{\Theta ((v_f p_s)^2-\epsilon^2)}
{\sqrt{(v_f p_s )^2-\epsilon ^2}}\right] \hat{{\bf p}}_s\,, \quad
|\epsilon|<\Delta_{0} \,.
\end{equation}

There is a simple relation between ${\bf j}(\epsilon , {\bf 0})$ and $N
(\epsilon ,{\bf 0})$ when $|\epsilon |<\Delta_{0}.$ In (\ref{jo}), the
delta function in the integrand of ${\bf j}(\epsilon , {\bf 0})$
effectively replaces $\hat{\bf k}$ by
$(\epsilon/v_f p_s)\hat{{\bf p}}_s$.
Taking this factor outside the integral leaves an integral
identical to that of the local density of states. Consequently,
\begin{equation}
\label{j}{\bf j}(\epsilon ,{\bf 0})=e\frac \epsilon { p_s }N
(\epsilon ,{\bf 0})\,\hat{{\bf p}}_s,\qquad |\epsilon |<\Delta_{0}. 
\end{equation}
Note that the contribution of negative energy (bound) states to the
total current density lies in the $-\hat{ {\bf p}}_s$ direction, i.e.
{\it opposite} to the externally imposed supercurrent. At zero
temperature the total current density originates from the bound states
having energies $-\Delta_0 < \epsilon < 0$,
\begin{equation}\label{j_bound}
{\bf j}_{bound}({\bf 0,}T=0)\,=\, \int_{-\Delta_{0}}^0\,{\bf j}(\epsilon
,{\bf 0})d\epsilon =-\frac{2eN_f\Delta_{0}}{L/\xi}v_f\text{\/}\hat{ 
{\bf p}}_s\,.
\end{equation}
The current density of an isolated vortex with $p_s=0$ vanishes at the
center of the vortex, {\it i.e.} ${\bf j}_3({\bf 0,}T)=0$. We can
combine the result in (\ref{j_bound}) with ${\bf j}_1$ given in
(\ref{j1}) to obtain the total current density at $T=0$:
\begin{equation}
\label{center}{\bf j}({\bf r=0,}T=0)=ev_f^2N_f{\bf p}_s-\frac{2N_fe\Delta
_{0}}{L/\xi}v_f\text{\/}\hat{{\bf p}}_s. 
\end{equation}
Thus, for sufficiently small {\bf p}$_s$ the bound-state
contribution dominates (\ref{center}) and ${\bf j}({\bf r=0,}T=0)$ will
point in the $-\hat{{\bf p}}_s$ direction.

\subsection{\label{div}Particle conservation}

For equilibrium conditions the divergence of the current density
vanishes.
From (\ref{j1}), ${\bf j}_1$ has a vanishing divergence, and for an
undisturbed vortex we have $\mbox{\boldmath $\nabla$}\cdot{\bf
j}_3({\bf r})=0$. Thus, $S({\bf r})= \mbox{\boldmath $\nabla$}\cdot{\bf
j}({\bf r})\equiv \mbox{\boldmath $\nabla$}\cdot{\bf j}_2({\bf r})$. At
$T=0$,
\begin{equation}
\label{so}S({\bf r})=4\pi ev_fN_f\int_0^{2\pi }\frac{d\varphi _k}{2\pi } 
\int_0^{-{\bf q\cdot}\hat{\bf k}}d\epsilon \,v_f\frac \partial {\partial \zeta
}\left( \langle \zeta |\delta (\epsilon -\hat{H}_A)|\zeta \rangle
_{1,1}\right) \,.
\end{equation}

In Eq. (\ref{andreevspectrum}) of appendix \ref{simplification} we show that
\begin{equation}\label{e}
v_f\frac \partial {\partial \zeta } \langle \zeta
|\delta (\epsilon -\hat{H}_A)|\zeta \rangle _{1,1}
=
\Delta_{0}\frac{F(r)}r\mbox{tr} 
\left[ (\zeta \hat\tau_{2}-\eta \hat\tau_{1})\langle \zeta |\delta (\epsilon - 
\hat{H}_A)|\zeta \rangle \right] 
\,,
\end{equation}
yielding 
\begin{equation}
\label{sso}S({\bf r})=4\pi ev_f\Delta_{0}N_f\int_0^{2\pi }\frac{ 
d\varphi _k}{2\pi }\int_0^{-{\bf q\cdot}\hat{\bf k}}d\epsilon \text{\/}\frac{ 
F(r)}r\mbox{Tr}\left[ (\zeta \hat\tau_{2}-\eta \hat\tau_{1})\langle \zeta
|\delta (\epsilon -\hat{H}_A)|\zeta \rangle \right] .
\end{equation}
Since $q\ll \Delta_{0}$, only the bound state, $\psi _0(\zeta ;\eta )$,
contributes to the $\epsilon $
integral. Thus,
\begin{equation}
\label{s}
S({\bf r})= 4\pi ev_fN_f\Delta_{0}\int_0^{2\pi }\frac{d\varphi _k}{ 
2\pi }\int_0^{-{\bf q\cdot}\hat{\bf k}}d\epsilon\, \text{\/}\delta (\epsilon
-\epsilon_0(\eta ))\,\psi _0^{\dagger }(\zeta ;\eta )\frac{F(r)}r(\hat\tau_{2}\zeta
-\hat\tau_{1}\eta )\psi _0(\zeta ;\eta )\,,
\end{equation}
where $\epsilon_0(\eta )$ is the bound-state energy for an impact
parameter $\eta$.  At small distances from the center of the vortex
($r\ll \xi$) $F(r)\approx r/\xi$, with $\xi\approx v_f/2\Delta_{0}$,
and the lowest energy bound state is then
\begin{equation}
\psi _0(\zeta ;\eta ) = \frac 1{\pi ^{1/4}}\frac 1{\sqrt{\xi}}\exp
\left( -\frac{\zeta ^2}{2\xi^2}\right) \left( 
\begin{array}{c}
\frac 1{
\sqrt{2}} \\ -\frac i{\sqrt{2}}
\end{array}
\right) \,,
\end{equation}
\begin{equation}
\label{eig}
\epsilon_0(\eta ) = -\Delta_{0}\frac \eta {\xi}.
\end{equation}
Substituting $\psi _0$ and $\epsilon_0$ into (\ref{s}) and setting $\exp (-\frac
12\zeta ^2/\xi^2)\approx 1$ yields 
\begin{equation}
\label{sfinal}
S({\bf r})  \approx   \frac 4{\sqrt{\pi }}N_f\Delta_{0}^2\frac{\frac{ 
{\bf q}}{\Delta_{0}}{\bf \cdot }\frac{{\bf r}}{\xi}}{\left| \frac
r{\xi}\hat{\varphi }-\frac{{\bf q}}{\Delta_{0}}\right| }\,,
\end{equation}
which is non-zero, indicating that the ansatz (\ref{delta}) is not
physically correct for any value of ${\bf p_s}$. This failure to
satisfy the conservation law is due to the lack of self-consistency of
the order parameter in Eq. (\ref{delta}) in the presence of the flow
field.  In the absence of pinning, the  vortex will move in response to
a flow field, even one of arbitrarily small strength. The results
in section II implicitly assume a {\sl pinned} vortex. Thus, there will
be distortion of the vortex away from its cylindrically symmetric
equilibrium form (\ref{delta}). In the following section we show that
the self-consistently determined vortex order parameter, which includes
the deformation by the flow field, restores the conservation law.

\section{Quasiclassical Results}

A versatile and efficient method for calculating local spectral
properties of superconductors is the quasiclassical theory of
superconductivity.\cite{eil68,lar69,eli72,lar76} This theory is well
adapted for  a numerical approach to microscopic problems in
superconductivity, such as the calculation of the structure and the
excitation spectrum of vortex cores for superconductors with isotropic,
anisotropic or unconventional order parameters. The quasiclassical
theory is the only theoretical formulation which can handle equally
well clean and dirty superconductors, as well as more complicated
geometries than that of an isolated vortex with cylindrical symmetry or
a perfect vortex lattice. It can be interpreted as the generalization
of Landau's theory of normal Fermi liquids to the superconducting
state.  The quasiclassical theory shares with Landau's theory the
semiclassical description of the orbital degrees of freedom of
quasiparticle excitations. On the other hand, the internal degrees of
freedom, {\it i.e.} the spin and the particle-hole degrees of freedom,
are described by quantum mechanics. Quantum mechanical coherence of
particle and hole excitations is  the basis of all superconducting
effects such as persistent supercurrents, flux quantization, Josephson
effects and Andreev reflection. Here we use the quasiclassical theory
for our investigations of the vortex core. Numerical work on vortices
in superconductors using the quasiclassical theory started with a
series of publications by Kramer, Pesch, and Watts-Tobin.
\cite{kra74,pes74,wat74} More recent work includes pinning of vortices
at small defects,\cite{thu84} vortices in superfluid $^3$He and other
systems with unconventional pairing,\cite{thu90,fog95} the excitation
spectrum  of bound quasiparticles,\cite{kle90,sch95} and the spectrum
of moving pancake vortices.\cite{hof95}

We use in this article the notation of Refs. \onlinecite{ser83,rai86,rai94}. 
The central objects of the quasiclassical theory of superconductors
in equilibrium are the quasiclassical propagators $\hat g^{R,A}({\bf
p}_f,{\bf r};\epsilon)$, which  are $2\times 2$ matrices in the
particle-hole index,
\begin{equation}\label{matrx}
  \hat g^{R,A}=
  \left(
  \begin{array}{cc}
  g^{R,A}({\bf p}_f,{\bf r};\epsilon)
  &f^{R,A}({\bf p}_f,{\bf r};\epsilon)
  \\
  \underline{f}^{R,A}({\bf p}_f,{\bf r};\epsilon)
 &\underline{g}^{R,A}({\bf p}_f,{\bf r};\epsilon)
  \end{array}\right)\ .
  \end{equation}
The variables $\epsilon$ and ${\bf p}_f$ stand for the energy of an
excitation and its momentum (on the Fermi surface). The momentum
variable reduces to ${\bf p}_f=p_f(\cos\varphi_k,\, \sin\varphi_k)$ for
an isotropic Fermi surface in two dimensions (see section II).  General
symmetries lead to the following fundamental relation between $\hat
g^R$ and $\hat g^A$,
\begin{equation}
\hat g^A\,=\,\hat\tau_3 \left(\hat g^R\right)^{\dagger}\hat\tau_3\, .
\end{equation}
We use, as described in Section II, 
the notation $\hat\tau_{1}$, $\hat\tau_{2}$, $\hat\tau_{3}$, for the three Pauli matrices in particle-hole space, and $\hat 1$ for the unit matrix.
The off-diagonal terms $f^{R,A}$ in Eq. (\ref{matrx})
are the pair amplitudes. They vanish in the normal state, and measure
the amount of particle-hole mixing in the superconducting state.
The diagonal elements of the propagators determine the density
of states,
\begin{equation}
N({\bf p}_f,{\bf r};\epsilon)=
N_f
{g^R({\bf p}_f,{\bf r};\epsilon)-g^A({\bf p}_f,{\bf r};\epsilon)\over -2\pi i}\, ,
\end{equation}
and the equilibrium current density. The most detailed information on
the current distribution is obtained from  the {\it spectral current
density},
%\begin{equation}
%{\bf j}({\bf p}_f,{\bf r};\epsilon)= e {\bf v}_f N_f
%{g^R({\bf p}_f,{\bf r};\epsilon)-
%g^A({\bf p}_f,{\bf r};\epsilon)\over -2\pi i}\, ,
%\end{equation}
\begin{equation}
{\bf j}({\bf p}_f,{\bf r};\epsilon)= e {\bf v}_f N_f\,
\left({\cal N}_{+}({\bf p}_f,{\bf r};\epsilon) - {\cal N}_{-}({\bf p}_f,{\bf r};\epsilon)\right)
\, ,
\end{equation}
where ${\cal N}_{\pm}({\bf p}_f,{\bf r};\epsilon)=N(\pm{\bf
p}_f,{\bf r};\epsilon)/N_f$ is the dimensionless density of states for co-moving
($+$) and counter-moving ($-$) excitations along the trajectory line
defined by ${\bf p}_f$, and ${\bf v}_f$ is the Fermi velocity at the
point ${\bf p}_f$ on the Fermi surface.  This spectral density measures
the contributions of quasiparticle states with energy $\epsilon$ and
momentum near the Fermi surface point ${\bf p}_f$ to the current
density at position ${\bf r}$. The full current density is obtained by
weighting the spectrally resolved current density by the occupation
probability of the quasiparticle states, then integrating over Fermi
momenta and energies. For equilibrium states,
\begin{equation}
{\bf j}({\bf r})=2\int d\epsilon\int d{\bf p}_f\
{\bf j}({\bf p}_f,{\bf r};\epsilon)
\left(f(\epsilon)-{1\over 2}
\right)\, ,
\end{equation}
where $f(\epsilon)$ is the Fermi distribution function. The symbol
$\int d{\bf p}_f$ denotes a weighted integral over the Fermi surface.
The weight at ${\bf p}_f$ is $\propto \mid{\bf v}_f\mid^{-1}$, and the
integral is normalized, $\int d{\bf p}_f\ 1\,=\, 1$. 
The spectral current density
is particularly well suited for our study of
the importance  of Andreev bound states for the current flow in a
vortex core. These bound states appear as delta functions in the
spectral current density at energies below the bulk energy
gap. The spectral weight of the delta function, combined with the
occupation of the bound state, determine its contribution to the total
current density. 

We calculate $\hat g^R({\bf p}_f,{\bf r};\epsilon)$ from
Eilenberger's transport equation\cite{eil68}
\begin{equation}\label{eilen}
\left[(\epsilon+{e\over c}{\bf v}_f\cdot {\bf A}({\bf r}))\hat\tau_3 
-\hat\Delta({\bf p}_f,{\bf r}),\, \hat g^R({\bf p}_f,{\bf r};\epsilon)\right]
+i\hbar{\bf v}_f\cdot\mbox{\boldmath $\nabla$}\hat g^R({\bf p}_f,{\bf r};\epsilon)=0\, ,
\end{equation}
supplemented by the condition of analyticity in the upper half of the
complex $\epsilon$-plane, and the normalization condition
\begin{equation}
\hat g^R({\bf p}_f,{\bf r};\epsilon)^{\, 2}=-\pi^2\hat 1\, .
\end{equation}
For a fixed Fermi momentum ${\bf p}_f$ this is a first order
differential equation along a straight-line classical trajectory in the
direction of the Fermi velocity ${\bf v}_f$. The propagator $\hat
g^R({\bf p}_f,{\bf r};\epsilon)$ at a chosen point of interest, ${\bf
r}$, is determined by the solution of (\ref{eilen}) along the
trajectory through ${\bf r}$ in the direction ${\bf v}_f$. Complete
information on the local physical properties at point  ${\bf r}$, such
as the current density, is obtained by sampling all trajectories
through ${\bf r}$. The propagator $\hat g^R({\bf p}_f,{\bf
r};\epsilon)$ is intimately related to the $2\times2$ density matrix of
the particle-hole degrees of freedom of a quasiparticle moving along
the classical trajectory specified by ${\bf p}_f$, ${\bf r}$. Thus,
$\hat g^R({\bf p}_f,{\bf r};\epsilon)$ describes the state of the
internal degrees of freedom of the excitation. The internal state, {\it
i.e.}  the amount of particle-hole mixing, may change along the
trajectory as a consequence of the off-diagonal pair potential, $\hat
\Delta({\bf p}_f,{\bf r})$, which acts as a driving term that `rotates'
the particle-hole pseudo spin. The pair potential couples particle and
hole excitations, and is the origin of particle-hole coherence. It
depends on the real space position, ${\bf r}$, and, for anisotropic
superconductors, on the Fermi surface position, ${\bf p}_f$,
\begin{equation}\label{matrxdelta}
\hat \Delta({\bf p}_f,{\bf r}) =
\left(
\begin{array}{cc}
 0 
&\Delta({\bf p}_f,{\bf r}) \\
-\Delta^{\ast}({\bf p}_f,{\bf r}) 
&0
\end{array}\right)\ .
\end{equation}
The pair potential must be calculated self-consistently from
the gap equation,
\begin{equation}\label{gapeq}
\Delta({\bf p}_f,{\bf r})=
\int d{\bf p}_f^{\ \prime}\, V({\bf p}_f,{\bf p}_f^{\ \prime})
\int {d\epsilon\over 2\pi}
\,{\cal I}m\,f^R({\bf p}_f^{\ \prime},{\bf r};\epsilon)\,
(1-2f(\epsilon))\, ,
\label{gapequation}
\end{equation}
where $V({\bf p}_f,{\bf p}_f')$ is the pairing interaction, which
determines the
orbital symmetry of the pair potential, its magnitude and $T_c$.

Our procedure for numerical calculation of the currents in the core of
2D pancake vortices is the following. We first solve self-consistently
the gap equation and Eilenberger's equation at Matsubara energies.
This allows us to determine the pair potential and the supercurrent
density.  We then insert the pair potential into Eilenberger's
differential equation at real energies, and obtain from its solution
the excitation spectrum:  the density of states and the spectral
current density. The differential equations are solved by standard
4$^{th}$ order Runge-Kutta and multiple shooting methods, and
self-consistency is achieved iteratively by using alternatively a
relaxation method and the M\"obius-Eschrig algorithm.\cite{esc89}

We consider three examples of pancake vortices: isolated, i)
singly-quantized and ii) doubly-quantized $s$-wave vortices, and iii) a
pinned $s$-wave vortex in the presence of a uniform transport
supercurrent.  We choose a temperature of $T=0.4\, T_c$, unless
otherwise noted, and assume $\kappa=\lambda/\xi\gg 1$. In this limit
the vector potential is essentially constant in the core region, and
can be neglected.

\subsection{Spectral current density of a singly-quantized s-wave vortex}

Figure 1 shows the amplitude of the order parameter of a
singly-quantized $s$-wave vortex. The amplitude is isotropic and
vanishes linearly in the core. The variation of the amplitude and phase
along two trajectories are also shown in Fig. 1. For trajectory $a$
passing through the center of the vortex, the phase changes
discontinuously and the amplitude vanishes linearly at the vortex
center.  For trajectory $b$, with impact parameter $\eta=3.0\xi_0$,
there is only a small change in the amplitude of $\Delta$. For
singly-quantized vortices the phase of the order parameter is the more
important factor determining the spectrum of bound states.

Figure 2 shows the angle-resolved local density of states for the  two
trajectories shown in Fig. 1. For trajectory (a) through the center of
the vortex, the spectrum shows a zero-energy bound state separated from
the continuum that begins at the bulk gap. The bound state results from
constructive interference of particle- and hole-like quasiparticles
that undergo Andreev reflections from the vortex order parameter.  This
bound state corresponds to the zero angular momentum bound state found
by Caroli, de Gennes and Matricon.\cite{car64,car65} A zero-energy
bound state is always present for trajectories in which the order
parameter is real (up to a constant phase factor) and has different
signs when going to $\pm\infty$ along the trajectory.\cite{ati75}

Bound states with non-zero energies are found for trajectories with a
nonzero impact parameter measured from the vortex center. These bound
states correspond to the spectrum of bound states with non-zero angular
momenta obtained by Caroli et al.\cite{car64} Figure 2b shows the
spectrum for a trajectory with an impact parameter of
$\eta=4.2\,\xi_{0}$ and ${\bf v}_f\cdot{\bf p}_s({\bf r})\ge 0$
measured at the point of closest approach to the vortex center.  The
bound state is shifted down in energy to $\epsilon/2\pi T_c \simeq
-0.22 $, and the continuum states are shifted and inhomogeneously
broadened by the Doppler energy, $\Delta\epsilon={\bf v}_f\cdot{\bf
p}_s({\bf r})$. The spectrum near the onset at point $1$ in Fig. 2b has
low weight and corresponds to the continuum edge at $\epsilon=\Delta$
far from the impact point, while the peak in the spectrum at point $2$
corresponds to the maximum Doppler shift, $\epsilon=\Delta+{\bf
v}_f\cdot{\bf p}_s({\bf R})$ at the impact point ${\bf R}$. Note the
development of the BCS coherence peak as the density of states is
sampled further from the vortex center.

\newpage

%%%%%%%%%%%%%%%%%%%%%%%%%% Fig. 1 %%%%%%%%%%%%%%%%%%%%%%%%%%%%%%%%%
\begin{multicols}{2}
\noindent\begin{figure}
\epsfxsize=\hsize\epsfbox{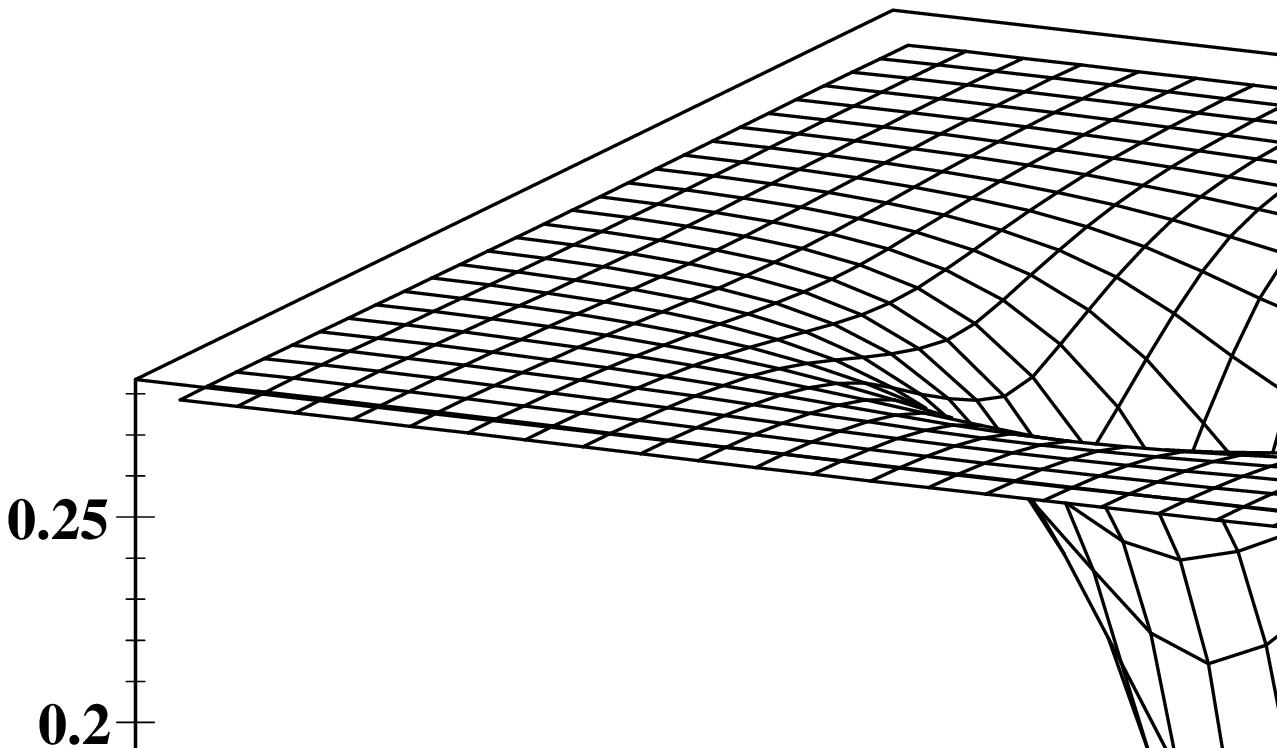}
\end{figure}

\begin{figure}
\epsfxsize=\hsize\epsfbox{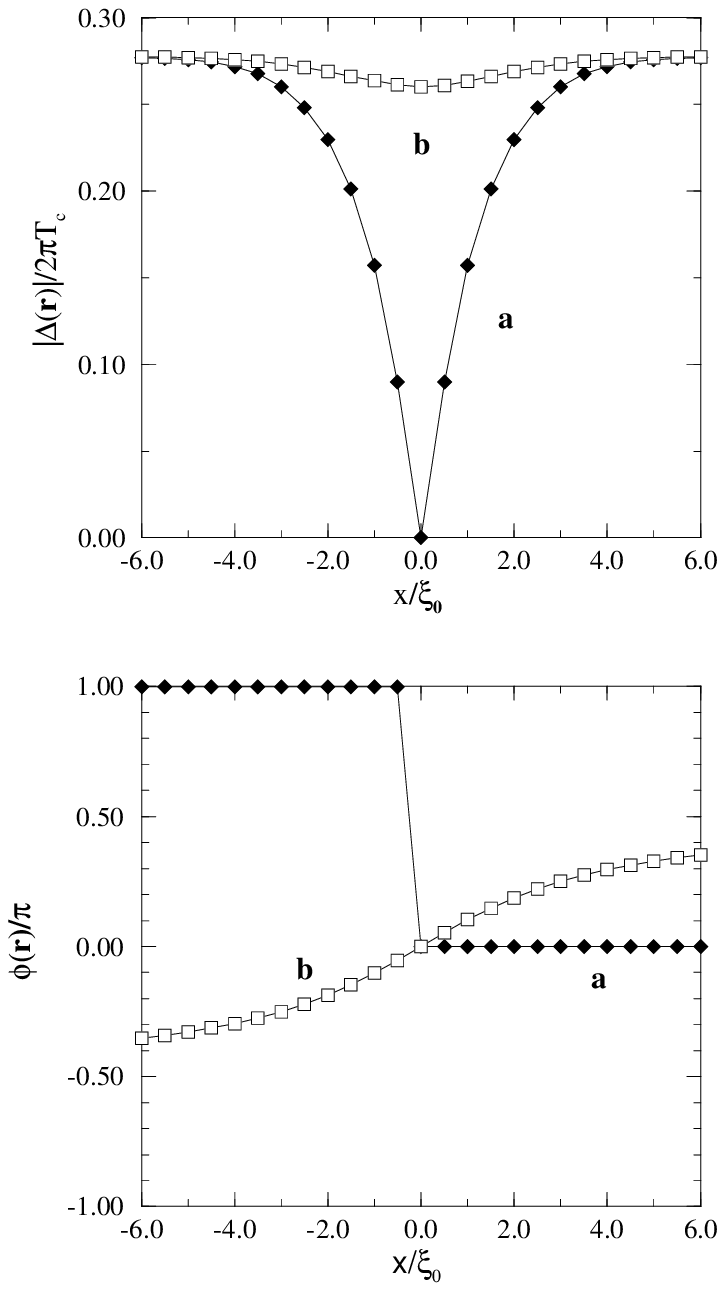}
\end{figure}
\end{multicols}
\begin{figure}
\begin{center}
\begin{minipage}{0.85\textwidth}
\caption{
The magnitude of the pair potential, $|\Delta({\bf r})|/2\pi T_c$, at
$T=0.4T_c$ for a singly quantized vortex in an s-wave superconductor
is shown in the 3D plot. The 2D plots show the order parameter
amplitude and phase of the order parameter along a trajectory (a)
passing through the center of the vortex, and (b) along a trajectory
with an impact parameter of $\eta=3.0\xi_0$. The order parameter is
real along trajectory (a) and the phase changes discontinuously by
$\pi$. Along trajectory (b) there is little change in amplitude, but a
substantial, continuous change of phase.}
\label{swave_OP-Phase}
\end{minipage}
\end{center}
\end{figure}
%%%%%%%%%%%%%%%%%%%%%%%%%%%%%%%%%%%%%%%%%%%%%%%%%%%%%%%%%%%%%%%%

The density of states of an $s$-wave  vortices has been investigated by
several authors.\cite{pes74,kle90,ull90}  Our emphasis is on the
importance of the Andreev bound states for the current distribution in
the vortex core. We show in Fig. 2c the spectral current density for
the trajectory with $\eta= 4.2\,\xi_0$ and ${\bf v}_f\cdot{\bf
p}_s({\bf r})\ge 0$. The net current carried by the states at the point
$\pm{\bf p}_f$ on the Fermi surface is obtained by weighting this
spectrum by the equilibrium distribution and integrating over all
energies. Thus, for $T\rightarrow 0$ only the negative energy states
contribute. It is clear from Fig. 2c that the current in the vicinity
of the vortex core is carried almost entirely by the bound states with
$-|\Delta|<\epsilon<0$.  The continuum states give almost no net
contribution to the current in the core. Figure 2d shows the spectral
current density of the set of bound states with trajectories ${\bf
v}_f=\pm v_f\hat{\bf y}$ as a function of the impact parameter $\eta$
for $0\le\eta\le 6\xi_0$. The peak at $\epsilon/2\pi T_c\simeq -0.027$
corresponds to the trajectory with impact parameter $\eta=0.2\xi_0$.
The bound state energy decreases with increasing distance from the
core. For small $\eta$ we obtain, $\epsilon_{0}(\eta)\simeq
-2(\eta/\xi_0)\Delta$, in reasonable agreement with the analytic
estimate in Eq. (\ref{eig}). As indicated in Fig. 2d the contribution
of the bound state to the current density decreases as the impact
parameter increases. However, even at a relatively large distance,
$\eta=6\xi_0$, the bound state still contributes significantly to the
circulating current density of the vortex.

The evolution of the bound state energy for small impact parameters can
be written in terms of the angular momentum of an excitation about the
vortex center, ${\cal L}_z=p_f\eta$; {\it i.e.}
$\epsilon_0(\eta)=-{\cal L}_z\,\omega_0$, where
$\hbar\omega_0=2\hbar\Delta/p_f\xi_0\ll\Delta$. This spectrum was
originally obtained by Caroli et al.\cite{car64} by solving the
Bogolyubov equations. In the Bogolyubov or Gor'kov formulation the
spectrum is discrete: ${\cal L}_z=(m+\case{1}{2})\,\hbar$ with
$m=\,\mbox{integer}$ and $\hbar\omega_0$ defining the level spacing of
the low-lying bound states in the core. The lowest energy bound state
in the core has a zero-point energy of
$\epsilon_0=\case{1}{2}\hbar\omega_0\simeq\Delta^2/E_f\ll\Delta$ which
is outside the resolution of the quasiclassical or the Andreev theory.
The discrete spectrum of the Bogolyubov theory corresponds to the
continuous Andreev spectrum in the limit where the level spacing is
small compared to all other relevant energy scales, {\it i.e.}
$\hbar\omega_0\ll k_BT\,,\,\hbar/\tau$, etc. This is generally an
excellent approximation in conventional type II superconductors. For the high $T_c$ cuprates the discrete level structure is expected to
play a more important role, particularly in the transverse response of
vortices in the ultra-clean limit, {\it i.e.} $\omega_0\gg 1/\tau$,
where $\tau$ is the mean scattering time.\cite{kop76,kop96}

%%%%%%%%%%%%%%%%%%%%%%%%% Figure 2 %%%%%%%%%%%%%%%%%%%%%%%%%%%%%%%%
\vspace{0.2cm}
\begin{figure}
\begin{center}
\begin{minipage}{0.6\textwidth}
\epsfxsize=\hsize
\epsfbox{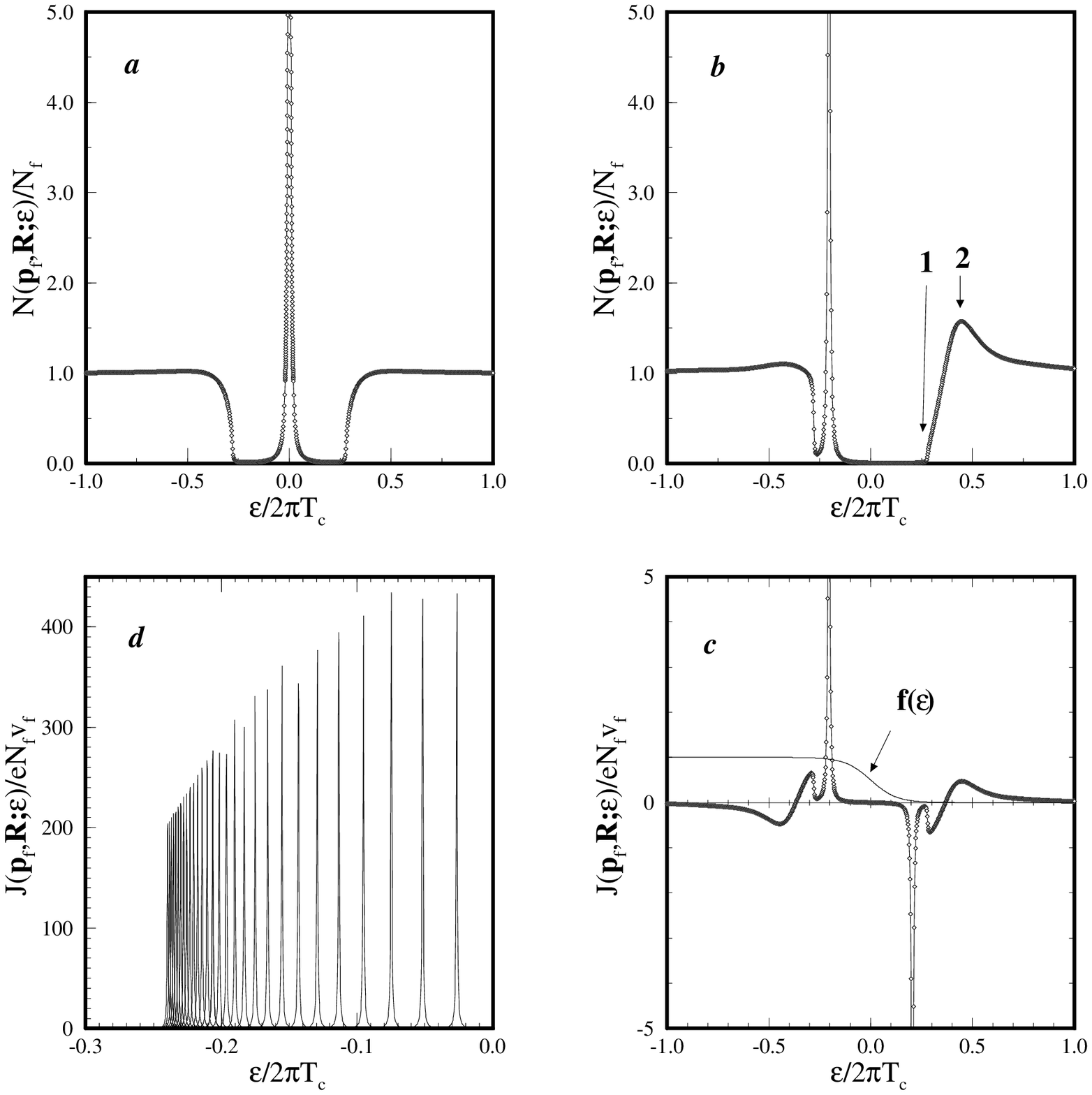}
\end{minipage}
\end{center}
\end{figure}
\begin{figure}
\begin{center}
\begin{minipage}{0.85\textwidth}
\caption{$a$) Local density of states at the center of a vortex for a
trajectory passing through the center of the core. The width of the
bound state is set at $\gamma/2\pi T_c=0.0004$, the continuum edge is
at $\epsilon=\pm\Delta$, and the temperature is $T=0.4T_c$. $b$) Local
density of states at ${\bf R}=(4.2,0)\xi_0$ for the trajectory ${\bf
v}_f=(0,1)v_f$. The bound state is shifted,
$\epsilon/2\pi T_c\simeq -0.22$, and the continuum states show the Doppler
broadening. $c$) The spectral current density for the same
position and direction as in $b$. The Fermi function for $T=0.4T_c$ is
also shown. Note that the current density is dominated by the negative energy
bound state. $d$) The spectral current density for a set of parallel
trajectories as a function of impact parameter for $0<\eta<6\xi_0$.
The spatial separation between neighboring trajectories is $0.2\xi_0$.
}
\label{Spectrum_swave}
\end{minipage}
\end{center}
\end{figure}
%%%%%%%%%%%%%%%%%%%%%%%%%%%%%%%%%%%%%%%%%%%%%%%%%%%%%%%%%%%%%%%%%%%

\subsection{Spectrum of a doubly-quantized s-wave vortex}

It is interesting to compare the single-quantum vortex with the axially
symmetric, $4\pi$ vortex, $\Delta({\bf r})=|\Delta({\bf
r})|\exp{2i\varphi}$.  The double quantum vortex has higher energy than
a pair of isolated single-quantum vortices; however, once created the
double-quantum vortex is metastable against dissociation into
singly-quantized vortices.  The amplitude of the order parameter for
the double-quantum vortex decreases as $|\Delta({\bf r})|\sim r^2$ for
$r < \xi_0$ as shown in Fig. 3a.

%%%%%%%%%%%%%%%%%%%%%%%%% Figure 3 %%%%%%%%%%%%%%%%%%%%%%%%%%%%%%%%
\vspace{0.2cm}
\begin{figure}
\begin{center}
\begin{minipage}{0.6\textwidth}
\epsfxsize=\hsize
\epsfbox{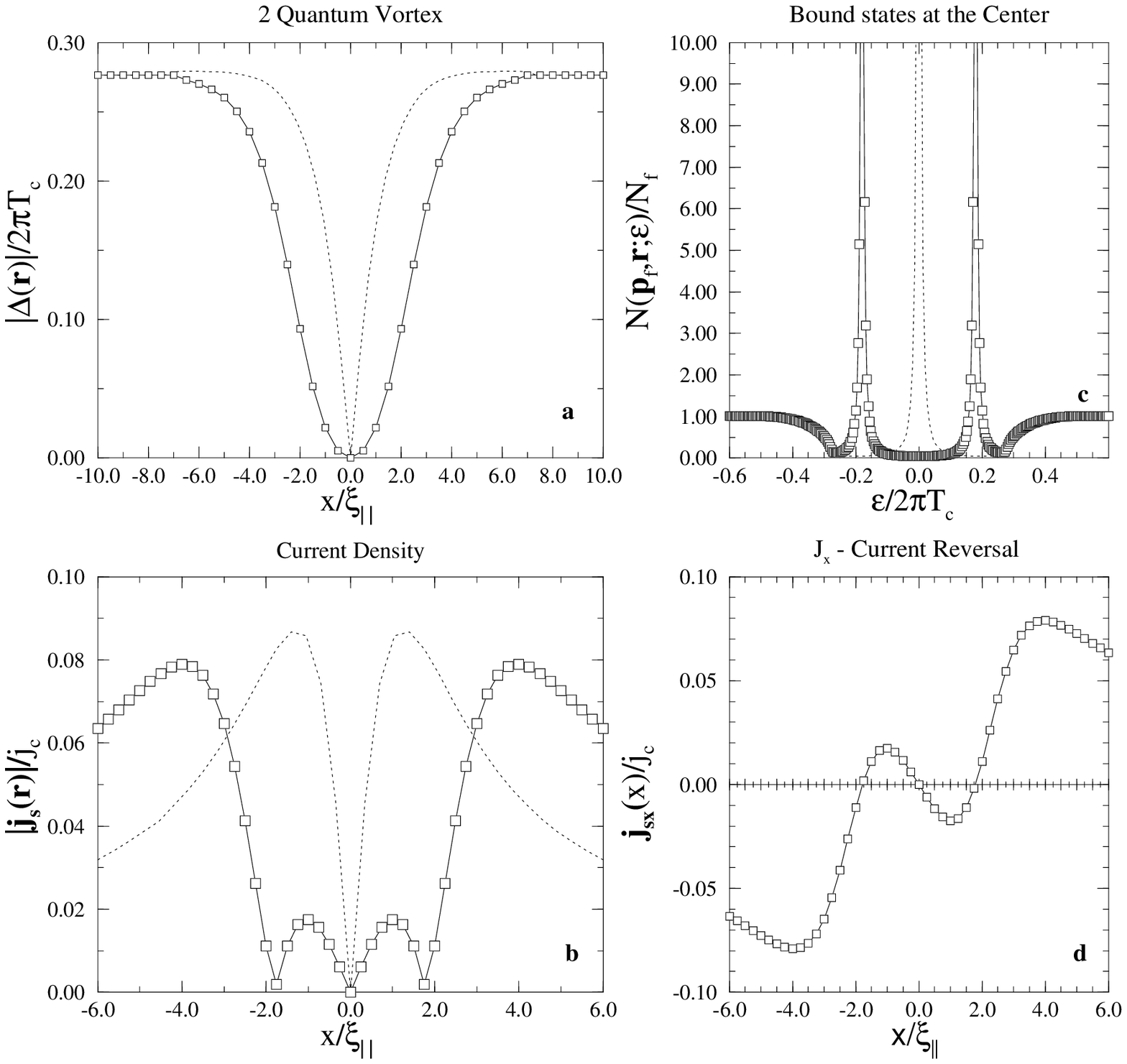}
\end{minipage}
\end{center}
\end{figure}
\begin{figure}
\begin{center}
\begin{minipage}{0.85\textwidth}
\caption{
$a$) The amplitude of the order parameter for a $4\pi$ vortex
at $T=0.4T_c$. Note the quadratic behavior for $r\ll\xi_0$. 
$b$) Local density of states at the center of the vortex for a
trajectory passing through the center of the core.  {\it Two} bound
states are present at energies, $\epsilon/2\pi T_c=\pm 0.18$.
$c$) The plot of $J_y(x,0)$ vs. $x$ shows a {\it reversal} of the
direction of the current for $x<1.9\xi_0$.
$d$) The magnitude of the current density for the $4\pi$ vortex.
The corresponding quantities for the $2\pi$ vortex are shown
for comparison (dotted curves).
}
\label{Doublequantum}
\end{minipage}
\end{center}
\end{figure}
%%%%%%%%%%%%%%%%%%%%%%%%%%%%%%%%%%%%%%%%%%%%%%%%%%%%%%%%%%%%%%%%%%%

In contrast to the $2\pi$ vortex there is no sign change of the order
parameter for trajectories passing through the center of the vortex
core. This difference has a profound effect on the spectrum of Andreev
bound states in the core. Fig. 3b shows the excitation spectrum of the
doubly-quantized vortex at the center of a trajectory passing through
the center of the vortex core. A symmetric spectrum of two bound states
at $\epsilon_{\pm}/2\pi T_c=\pm 0.18$ are separated from the
continuum.  Figure 3 also shows the current density of the
doubly-quantized vortex.  The remarkable feature is the {\it reversal}
of the current direction in the core, i.e. for $r\lesssim 2\xi_0$ (see
Fig. 3c and 3d). This current anomaly is associated with the appearance
of a counter-moving Andreev bound state below the Fermi level
($\epsilon=0$). The evolution of the spectral current density is shown
in Fig. 4. The trajectories are parallel to $\hat{\bf y}$ and the
spectral current density is shown as a function of the impact
parameter. At distances greater than $x\simeq 2\xi_0$ two bound states
lie below zero energy, and both states are co-moving with the
circulating phase gradient, ${\bf p}_s$. As the vortex core is
approached the co-moving bound state nearest the Fermi level moves to
higher energy, and a counter-moving bound state above the Fermi level
(not shown) moves to lower energy. These two states cross the Fermi
energy ($\epsilon=0$) at approximately $x=2\xi_0$, leading to a
reversal of the integrated current density inside the core. The
cumulative current density for each trajectory is shown as the thick
solid line in each panel of Fig. 4.

%%%%%%%%%%%%%%%%%%%%%%%%% Figure 4 %%%%%%%%%%%%%%%%%%%%%%%%%%%%%%%%
\vspace{0.2cm}
\begin{figure}
\begin{center}
\begin{minipage}{0.6\textwidth}
\epsfxsize=\hsize
\epsfbox{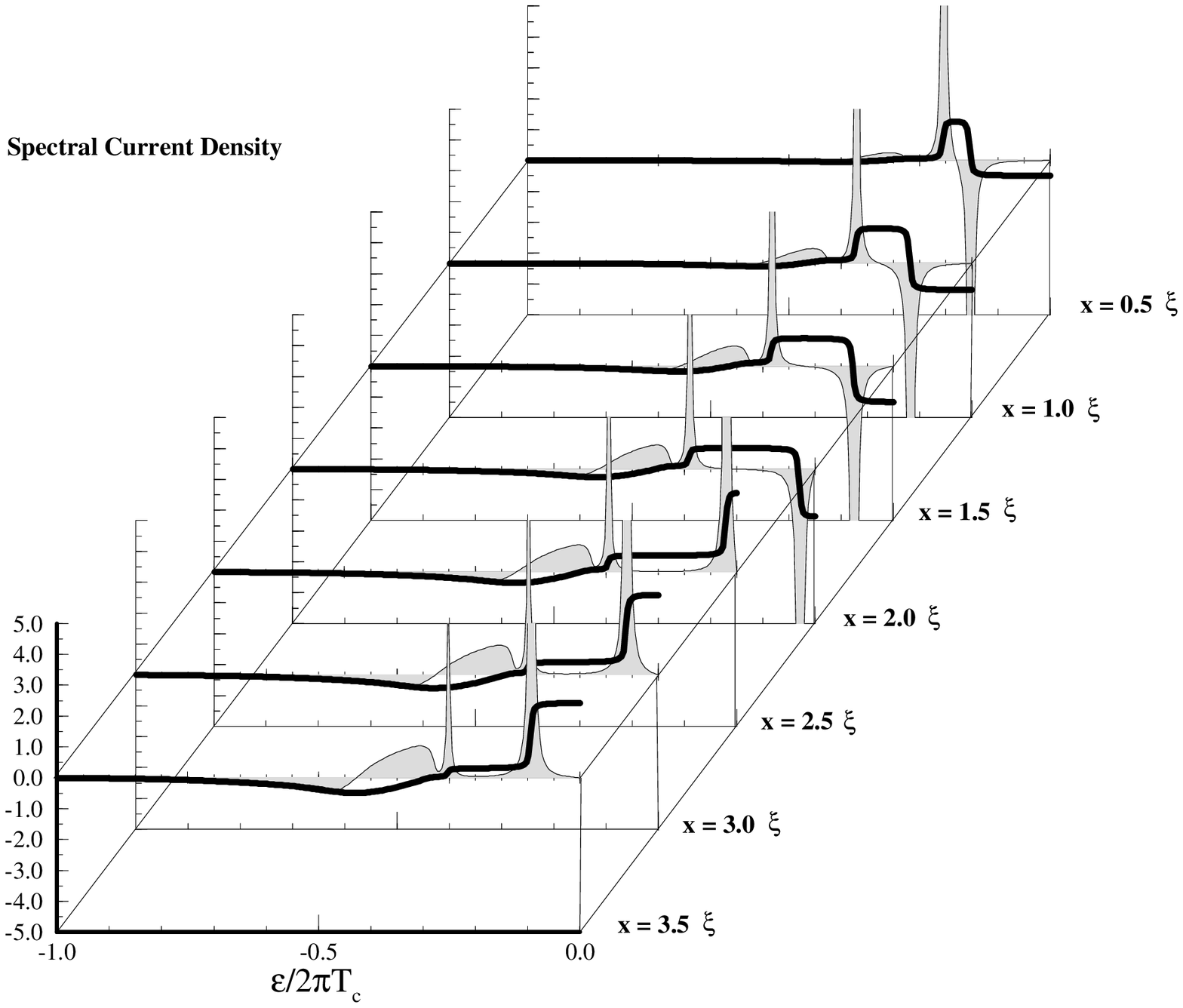}
\end{minipage}
\end{center}
\end{figure}
\begin{figure}
\begin{center}
\begin{minipage}{0.85\textwidth}
\caption{
The spectral current density of the $4\pi$ vortex for impact
parameters, $x=0.5\xi_0, ... , 3.5\xi_0$. The cumulative spectral
weight is shown as the thick solid line in each panel. Note the
appearance of the {\it counter-moving} bound state at $x=2.0\xi_0$ and the
corresponding reversal in the integrated spectral weight for
$x<2\xi_0$.
}
\label{JDOS_double}
\end{minipage}
\end{center}
\end{figure}
%%%%%%%%%%%%%%%%%%%%%%%%%%%%%%%%%%%%%%%%%%%%%%%%%%%%%%%%%%%%%%%%%%%

\subsection{Spectrum of a pinned s-wave vortex in a transport current}

Finally, consider the current and excitation spectrum of a $2\pi$ vortex in
the presence of a uniform supercurrent ${\bf j}_{tr}=j_{tr}\hat{\bf
x}$. In the absence of pinning the vortex will move in the direction
($-\hat{\bf y}$) in order to reduce the kinetic energy. Thus, in order
to investigate the excitation spectrum in the presence of a transport
current we must pin the vortex to the lattice. Our model for the
pinning center is a normal metal inclusion where the pairing
interaction (or the local $T_c$) vanishes.

%%%%%%%%%%%%%%%%%%%%%%%%% Figure 5 %%%%%%%%%%%%%%%%%%%%%%%%%%%%%%%%
\vspace{0.2cm}
\begin{figure}
\begin{center}
\begin{minipage}{0.6\textwidth}
\epsfxsize=\hsize
\epsfbox{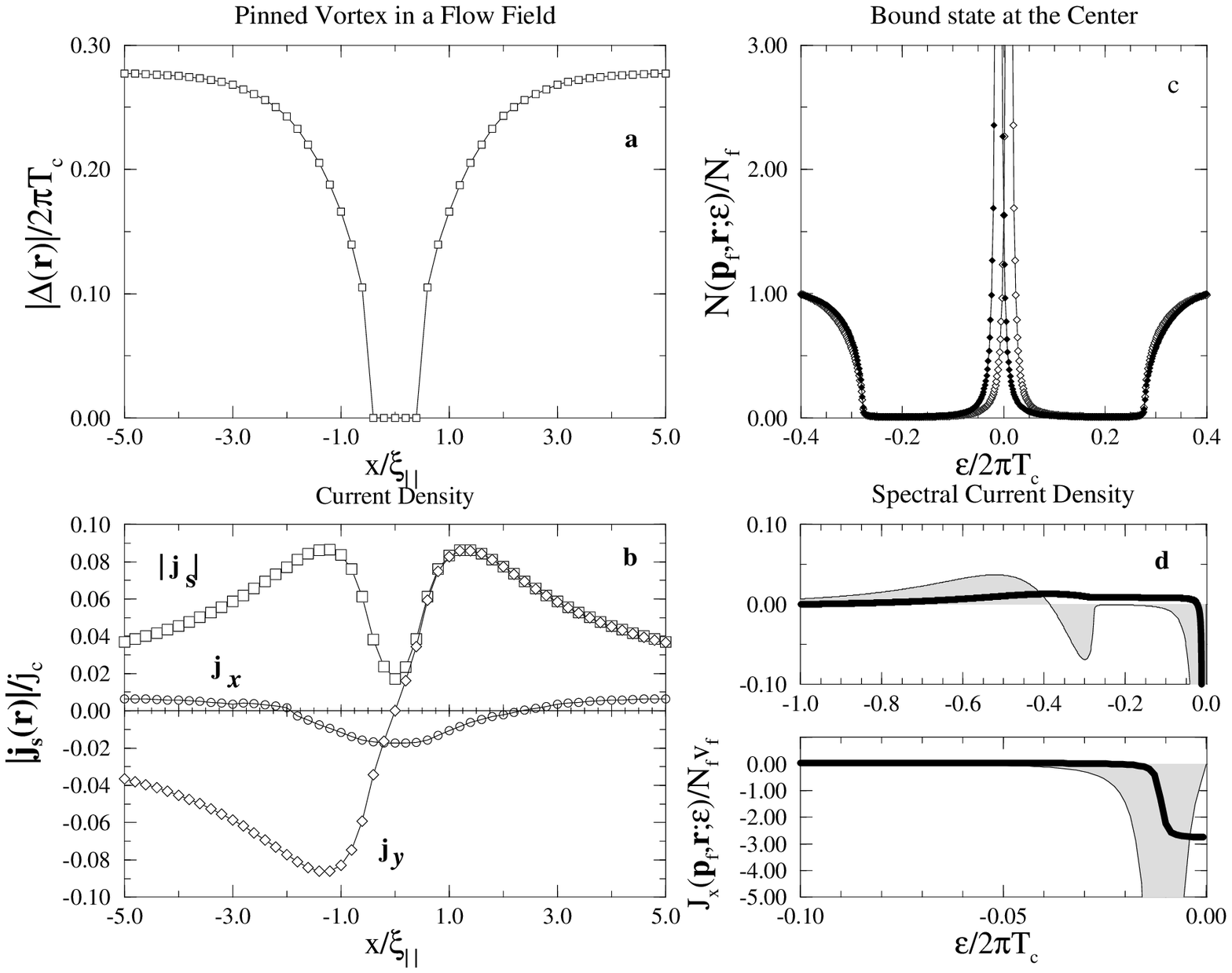}
\end{minipage}
\end{center}
\end{figure}
\begin{figure}
\begin{center}
\begin{minipage}{0.85\textwidth}
\caption{
$a$) The amplitude of the order parameter for a pinned $2\pi$ vortex at
$T=0.4T_c$. The normal inclusion has a diameter of $0.4\xi_0$, and the
imposed transport current corresponds to ${\bf
p}_s=0.02\xi_0^{-1}(1,0)$. The dotted curve corresponds to
$|\Delta(x)|$ in the absence of the normal inclusion.  $b$) Current
density of the pinned vortex for a trajectory passing through the
center of the core.  $c$) The bound state spectrum at the center of the
core of the pinned vortex in a uniform flow field. The two nearly
zero-energy bound states correspond to co-moving and counter-moving
trajectories.  $d$) The spectral current density for a trajectory
through the core. The negative energy counter-moving bound state
carries the backflow current in the core. The thick line is the
cumulative spectral weight for the current density. Note the scale
changes for the two panels.
}
\label{Pinnedvortex}
\end{minipage}
\end{center}
\end{figure}
%%%%%%%%%%%%%%%%%%%%%%%%%%%%%%%%%%%%%%%%%%%%%%%%%%%%%%%%%%%%%%%%%%%
Fig. 5a shows the order
parameter of a pinned vortex for an $s$-wave superconductor and a
pinning center with a radius of $0.4\xi_0$. In the presence of a
transport current the amplitude of the order parameter deforms; it is
suppressed on the high current side of the vortex as shown in Fig. 5a.
For relatively small transport currents, e.g. ${\bf
p}_s=0.02\xi_0^{-1}(1,0)$, the center for the phase winding lies within
the normal inclusion. However, as shown in Fig.  5b, the vortex current
no longer vanishes at the center of the vortex core; there is
substantial current {\it through} the vortex core region, including the
normal inclusion. The current density inside the normal inclusion is
carried by the Andreev bound states and is a consequence of the
proximity effect. The bound state spectrum at the center of the vortex
is shown for a trajectory parallel to the transport current and passing
through the vortex center. The negative energy bound state carries the
transport current inside the normal inclusion. Fig. 5d shows the
spectral current density measured at the center of the normal inclusion
for the trajectories with ${\bf v}_f\parallel {\bf p}_s$.  Note that
the bound state dominates the current and that this current is opposite
to the applied transport current. This result was also obtained in
section \ref{analyt}, without taking into account the distortion of the vortex
core order parameter. This led to a violation of charge conservation in
the core. Our numerical calculation shows that the main features of the
analytic model for the bound state spectrum and the self-consistent
determination of the order parameter for the pinned vortex in the
presence of a transport current guarantees that charge is conserved.
%%As a check of our numerics we evaluated $\mbox{\boldmath $\nabla$}\cdot{\bf j}$
%%and found $\xi_0\mid\mbox{\boldmath $\nabla$}\cdot{\bf j}/j_c < 10^{-5}$
%%everywhere.

\section*{Conclusions}

We have discussed the current carried by the excitations of s-wave
vortices in clean layered superconductors. The {\it spectral current
density} was introduced in order to identify the excitations that
determine the transport and circulating currents of a vortex. The bound
states of the vortex carry most of the current in the vicinity of the
core, including transport currents that flow through the core of a
pinned vortex. Far from the vortex core currents are carried primarily
by the bound-pair continuum that forms the condensate. For currents
flowing through a pinned vortex, current conservation is maintained by
``spectral transfer'' of the current carried by the Andreev bound
states to the continuum states outside the core. A novel example of the
evolution of the spectral current density is provided by the double
quantum vortex which shows the connection between the spectrum of bound
states and the symmetry or topology of the order parameter. At low
temperatures ($T=0.4 T_c$) the double quantum vortex exhibits a
`current reversal' relative to the asymptotic direction of the
circulation. The counter-circulating current in the core is due to a
counter-moving bound state that appears below the Fermi level and
dominates the current for distances of order $0 < r \lesssim 2\xi_0$.
At high temperatures, $T\rightarrow T_c$, this counter-moving bound
state is thermally depopulated with the result that the current
reversal in the core disappears in the Ginzburg-Landau limit. In
summary, we find that the Andreev bound states dominate the current of
vortices on the scale of a few coherence lengths. The nonequilibrium
properties of vortices on this scale are expected to be dominated by
the spectral evolution and dynamics of these bound states.

\section*{ACKNOWLEDGEMENTS}

This work was initiated when the authors were participants at a
workshop at the Institute for Scientific Interchange, Villa Gualino,
Torino. The research of DW was supported in part by the Engineering and
Physical Sciences Research Council, while that of DR and JAS was
supported in part by NSF grant DMR 91-20000 through the Science and
Technology Center for Superconductivity, the Max Planck Gesellschaft
and the Alexander von Humboldt Stiftung. We also thank Dr. M. J. Graf
for his comments on the manuscript.

\newpage

\section{Appendix}

\subsection{\label{simplification}A matrix element}

In this appendix we derive a form for a matrix element used in section \ref
{div}. The matrix element in question is $\langle \zeta |[i\/v_f\/\hat{p}_\zeta
,\delta (\epsilon -\hat{H}_A)]|\zeta \rangle _{1,1}$ where we write 
\begin{equation}
\label{ham}\hat{H}_A=v_f\/\hat{p}_\zeta \hat\tau_{3}+\hat{\Delta }%
,\qquad \hat{\Delta }\equiv \Delta _{0}\frac{F(\hat{r})}{\hat{%
r}}(\hat\tau_{1}\hat{\zeta }+\hat\tau_{2}\eta ).
\end{equation}
We have 
\begin{equation}
\lbrack i\/v_f\/\hat{p}_\zeta ,\delta (\epsilon -\hat{H}_A)] = 
i\left( \hat\tau_{3}(
\hat{H}_A-\hat{\Delta })\delta (\epsilon -\hat{H}_A)-\delta
(\epsilon -\hat{H}_A)(\hat{H}_A-\hat{\Delta })\hat\tau_{3}\right) 
\end{equation}
\begin{equation}
\qquad\qquad = i[\hat\tau_{3}\epsilon ,\delta (\epsilon -\hat{H}_A)]-i\{\hat\tau_{3}%
\hat{\Delta },\delta (\epsilon -\hat{H}_A)\}\,.
\end{equation}
Then 
\begin{equation}
\langle \zeta |[i\/v_f\/\hat{p}_\zeta ,\delta (\epsilon -\hat{H}%
_A)]|\zeta \rangle _{1,1}  =  i
\mbox{Tr}\left[ \frac 12(1+\hat\tau_{3})\langle 
\zeta |([\hat\tau_{3}\epsilon
,\delta (\epsilon -\hat{H}_A)]- \{\hat\tau_{3}
\hat{\Delta },\delta (\epsilon -\hat{H}_A)\})|\zeta \rangle \right] 
\end{equation}
\begin{equation}
\quad\qquad =  -\mbox{Tr}\left[ i\text{\/}\langle \zeta |\hat\tau_{3}\hat{\Delta }%
\delta (\epsilon -\hat{H}_A)|\zeta \rangle \right] \,.
\end{equation}
Substituting the explicit form for $\hat{\Delta }$ yields the relation
\begin{equation}\label{andreevspectrum}
\langle \zeta |[i\/v_f\/\hat{p}_\zeta ,\delta (\epsilon -\hat{H}%
_A)]|\zeta \rangle _{1,1}=\Delta _{0}\frac{F(r)}r\mbox{Tr}\left[
(\zeta \hat\tau_{2}-\eta \hat\tau_{1})\langle \zeta |\delta (\epsilon -\hat{H}%
_A)|\zeta \rangle \right] .
\end{equation}

\subsection{\label{far}Approximation of bound states at large distances}

Equation (\ref{joft}) gives the current density in terms of the Andreev
Hamiltonian (\ref{andreev}) whose eigenvalue equation reads

\begin{equation}
\label{eveqn}\left[ -iv_f\/\partial _\zeta \hat\tau_{3}+\Delta _{0}\frac{F(%
\sqrt{\zeta ^2+\eta ^2})}{\sqrt{\zeta ^2+\eta ^2}}(\hat\tau_{1}\zeta +\sigma
^2\eta )\right] \psi (\zeta )=E\psi (\zeta ).
\end{equation}
The parameter $\eta $
appearing in the above equation has the semiclassical interpretation as a
c-number impact parameter. In this appendix we present approximations to the
above equation for large values of the impact parameter, $\eta $.

For $|\eta |\gg \xi _0$ we are justified to replace $F(\sqrt{\zeta ^2+\eta ^2%
})$ by its asymptotic value of unity. Furthermore, making the somewhat crude
approximation 
\begin{equation}
\frac{(\hat\tau_{1}\zeta +\hat\tau_{2}\eta )}{\sqrt{\zeta ^2+\eta ^2}}\rightarrow
\hat\tau_{1}\frac \zeta {|\eta |}+\hat\tau_{2}\mbox{sign}(\eta )
\end{equation}
yields for the bound state 
\begin{equation}
\psi _0(\zeta )\simeq \text{constant}\times \exp \left( -\frac{\Delta _{0}%
}{2v_f|\eta |}\zeta ^2\right) \left( 
\begin{array}{c}
\frac 1{
\sqrt{2}} \\ \frac{-i}{\sqrt{2}}
\end{array}
\right) ,\qquad E_0\simeq -\Delta _{0}\mbox{sign}(\eta ).
\end{equation}
This  indicates that at large impact parameters, the bound states of
the Andreev equation are found very close to the threshold of the continuum.

When required, more refined estimates may be obtained by writing 
\begin{equation}
\frac{(\hat\tau_{1}\zeta +\hat\tau_{2}\eta )}{\sqrt{\zeta ^2+\eta ^2}}=\tau_1
\exp \left[ i\arctan (\eta /\zeta )\hat\tau_{3}\right] 
\end{equation}
and then performing a unitary transformation $\hat
H_A\rightarrow \hat U\hat H_A\hat U^{-1}$ , $%
\psi \rightarrow \widetilde{\psi }\equiv \hat U\psi $,  with $\hat 
U=\exp \left[ \frac
i2\arctan (\eta /\zeta )\hat\tau_{3}\right] $ to remove
the phase from the order parameter.   The transformed Andreev  equation is
\begin{equation}
\left[ -iv_f\/\partial _\zeta \hat\tau_{3}+\frac{v_f}2\frac \eta {\zeta ^2+\eta
^2}+\Delta _{0}\hat\tau_{1}\right] \widetilde{\psi }(\zeta )=E\widetilde{\psi 
}(\zeta )\,,
\end{equation}
This is  a one dimensional Dirac equation with a weak  scalar
potential, which has weakly bound states with energies near $\pm \Delta
_{0}$. A ``non-relativistic''
 treatment is appropriate in this case   and we approximate the Dirac
equation by a Schr\"odinger equation. For example, for $\eta <0$ we
write
\begin{equation}
\widetilde{\psi }=\psi _L\left( 
\begin{array}{c}
1 \\ 
1
\end{array}
\right) +\psi _S\left( 
\begin{array}{c}
1 \\ 
-1
\end{array}
\right) 
\end{equation}
with $\psi _{L,S}$ scalars. Straightforward manipulations indicate that $%
\psi _L$ approximately obeys the Schr\"odinger equation
\begin{equation}
\left[ -\frac{v_f^2}{2\Delta _{0}}\frac{\partial ^2}{\partial \zeta ^2}-%
\frac{v_f}2\frac{|\eta |}{\zeta ^2+\eta ^2}\right] \psi _L(\zeta )=(E-\Delta
_{0})\psi _L(\zeta ).
\end{equation}
All the machinery of Schr\"odinger theory may be used on this equation
to estimate e.g. the bound states. We can put  a lower limit on the
bound state energy. This may be obtained from the fact that the
eigenvalues of the Schr\"odinger operator are $\geq V_{\min }$, the
minimum of the potential. Thus, $E-\Delta _{0}\geq \mbox{Min}_\eta
\left[ - \frac{v_f}2\frac{|\eta |}{\zeta ^2+\eta ^2}\right] $, i.e.
\begin{equation}
E\geq \Delta _{0}-\frac{v_f}2\frac 1{|\eta |},\qquad \eta <0.
\end{equation}


\begin{thebibliography}{10}
\bibitem{f1}
We consider here only
superconductors with a spin-singlet order parameter, and neglect  the
(typically) small Pauli paramagnetism.
\bibitem{f3}
Note that the order parameter in (\ref{delta0})
and (\ref{delta}) depends on $\hat{r}=\sqrt{\hat{\zeta }^2+\eta ^2}$.
\end{thebibliography}
\end{document}